\documentclass[prl,aps,twocolumn]{revtex4}
\usepackage{amsmath,graphicx,epsfig}

\usepackage{epstopdf}
\usepackage{euscript}
\usepackage{amsfonts}
\usepackage{amssymb}
\usepackage{float}

\def\prn#1{{\left(#1\right)}}

\def\bra#1{{\langle#1|}}

\def\cg(#1,#2)(#3,#4)(#5,#6){\bra{#1,#2,#3,#4}#5,#6\rangle}

\def\ts#1{{_{\mbox{\scriptsize #1}}}}

\def\threej(#1,#2)(#3,#4)(#5,#6){\begin{pmatrix}#1&#3&#5\\#2&#4&#6\end{pmatrix}}
\def\sixj(#1,#2,#3)(#4,#5,#6){\begin{Bmatrix}#1&#2&#3\\#4&#5&#6\end{Bmatrix}}
\def\ninej(#1,#2,#3)(#4,#5,#6)(#7,#8,#9){\begin{Bmatrix}#1&#2&#3\\#4&#5&#6\\#7&#8&#9\end{Bmatrix}}

\def\sH{{\ensuremath{\EuScript H}}}

\def\mb{\mathbf}
\def\bs{\boldsymbol}

\newlength{\defbaselineskip}
\newcommand{\setlinespacing}[1]%
           {\setlength{\baselineskip}{#1 \defbaselineskip}}

\begin{document}

\title{Magnetic shielding and exotic spin-dependent interactions} 

\author{D. F. Jackson Kimball}
\email{derek.jacksonkimball@csueastbay.edu}
\affiliation{Department of Physics, California State University -- East Bay, Hayward, California 94542-3084, USA}

\author{J. Dudley}
\affiliation{Department of Physics, California State University -- East Bay, Hayward, California 94542-3084, USA}

\author{Y. Li}
\affiliation{Department of Physics, California State University -- East Bay, Hayward, California 94542-3084, USA}

\author{S. Thulasi}
\affiliation{Department of Physics, California State University -- East Bay, Hayward, California 94542-3084, USA}

\author{S. Pustelny}
\affiliation{Institute of Physics, Jagiellonian University, Lojasiewicza 11, 30-348, Krak\'ow, Poland}

\author{D.~Budker}
\affiliation{Helmholtz Institute Mainz, Johannes Gutenberg University, 55099 Mainz, Germany}
\affiliation{Department of Physics, University of California at Berkeley, Berkeley, California 94720-7300, USA}
\affiliation{Nuclear Science Division, Lawrence Berkeley National Laboratory, Berkeley, California 94720, USA}

\author{M. Zolotorev}
\affiliation{Center for Beam Physics, Lawrence Berkeley National Laboratory, Berkeley, California 94720, USA}

\date{\today}



\begin{abstract}
Experiments searching for exotic spin-dependent interactions typically employ magnetic shielding between the source of the exotic field and the interrogated spins. We explore the question of what effect magnetic shielding has on detectable signals induced by exotic fields. Our general conclusion is that for common experimental geometries and conditions, magnetic shields should not significantly reduce sensitivity to exotic spin-dependent interactions, especially when the technique of comagnetometry is used. However, exotic fields that couple to electron spin can induce magnetic fields in the interior of shields made of a soft ferro- or ferrimagnetic material. This induced magnetic field must be taken into account in the interpretation of experiments searching for new spin-dependent interactions and raises the possibility of using a flux concentrator inside magnetic shields to amplify exotic spin-dependent signals.
\end{abstract}



\maketitle

\section{Introduction}

With the discovery of the Higgs boson \cite{ATL12,CMS12}, the last major fundamental prediction of the Standard Model has been confirmed.  Nonetheless, there remain several fundamental mysteries in modern physics that defy explanation, such as the source of additional CP-violation required to explain baryogenesis \cite{Sak65,Gav94,Hue95}, the nature of dark matter \cite{Ber05} and dark energy \cite{Per99,Rie98}, the vacuum energy density catastrophe \cite{Wei89,Adl95}, and the hierarchy problem \cite{Ark98,Gra15}. To date no prediction of a new particle theory extending beyond the Standard Model has been unambiguously confirmed experimentally, and thus, in some sense, particle physics has entered a speculative era where most likely many possibilities must be explored before solutions to these fundamental mysteries are found. One such possibility is that the new physics required to explain baryogenesis, dark matter, or dark energy will be in the form of heretofore undiscovered fundamental forces or fields, and such exotic fields may interact with the intrinsic spins of elementary particles \cite{Moo84,Dob06,Fla09,Gra13}. There are several recent and ongoing laboratory experiments searching for such exotic spin-dependent couplings (see, for example, Refs.~\cite{Hun13,Hec08,Hec13,Ter15,Vas09,Kot15,Led13,Smi11,Kim13,Tul13,Bul13,Pos13,Pus13,Bud14,Arv14}, and also Ref.~\cite{Ant11} and Chapter 18 of Ref.~\cite{Bud13book} for reviews).

A typical laboratory experiment \cite{Ant11,Bud13book} searching for an anomalous spin-dependent interaction involves measuring the torque generated by an exotic field on an ensemble of spin-polarized particles (for example, by measuring the spin precession frequency of atoms in a vapor via optical pumping and probing \cite{Kim13} or measuring the rotation of a spin-polarized torsion pendulum \cite{Hec13,Ter15}). Usually the dominant torque on a spin comes from the interaction of a particle's magnetic moment $\bs{\mu}$ with the ambient magnetic field $\mb{B}$. To reduce noise and control systematic effects, the ensemble of spin-polarized particles is nearly always contained within a magnetic shield system that reduces the influence of background magnetic fields. Furthermore, the majority of recent experiments employ the technique of comagnetometry \cite{Bud13book}, where the torques on different species inhabiting the same volume are simultaneously measured.

In most experiments searching for anomalous spin-dependent interactions, the source of the exotic field is located outside the magnetic shield system. Any such experiment must answer the basic question: what is the effect of the magnetic shield system on the signal detected by the spin-polarized ensemble? It turns out that there are conflicting answers to this question in the literature: in Refs.~\cite{Vor88,Bob91}, it is argued that soft ferromagnets (common materials used in magnetic shields) are affected by exotic spin-dependent interactions; while in Refs.~\cite{Rit90,Pan92,Chu93}, it is argued that the response of soft ferromagnets to exotic interactions may be significantly reduced if the energy splitting between electron spin states caused by the exotic interaction is less than some threshold value. In the present paper we consider this question by analyzing the physical processes that lead to the shielding effect for different magnetic shield materials.

For soft ferromagnets, the effect of an exotic field that couples primarily to electron spin is essentially identical to that of a weak magnetic field, in the sense that both produce a torque on electron spins (whose dynamics are responsible for the shielding effect). One of the central questions in the case of soft ferromagnetic materials, as noted in the literature cited above \cite{Vor88,Bob91,Rit90,Pan92,Chu93}, is whether the permeability of soft ferromagnets depends on the external field strength in the small-field limit.  We address this question both theoretically, by considering the relevant physical processes leading to the shielding effect in soft ferromagnets, and experimentally, by carrying out measurements of the shielding effect for a mu-metal shield as a function of the applied external field $\mb{B}\ts{ext}$ in the small-field limit ($10^{-6}~{\rm G} \lesssim |\mb{B}\ts{ext}| \lesssim 10^{-3}~{\rm G}$) using an optical atomic magnetometer \cite{Kim09}. We demonstrate that in this field range the shielding effect is independent of magnetic field strength. This field range is of particular interest since it is the usual range of the ambient magnetic fields acting on the innermost shielding layer in experiments. Exotic fields would, in principle, generate an additional torque on electron spins in the shielding material on top of the torque from the ambient magnetic field.

In the next section, we consider the general features of the interaction of an exotic field coupling to intrinsic spin with magnetic shielding.

\section{Effect of exotic spin-dependent interactions on magnetic shielding}

Exotic spin-dependent couplings are generated in a wide variety of theories postulating new physics beyond the Standard Model: for example, theories incorporating new scalar/pseudoscalar or axial vector interactions \cite{Moo84,Dob06,Fla09,Gra13}, long-range torsion gravity \cite{Nev80,Nev82,Car94}, violation of Lorentz and CPT symmetries \cite{Kos11}, spontaneous breaking of Lorentz symmetry \cite{Ark05}, unparticles \cite{Geo07,Lia07}, and so on. In order for such interactions to have evaded experimental detection to this point, it must be the case that the coupling strength is much weaker than magnetic couplings or that the interaction range is shorter than atomic distance scales \cite{Kot15,Led13}. This work concentrates on the former possibility since we are interested in cases where magnetic shielding may be interposed between the source of the exotic field and the spins used to detect the interaction. Furthermore, such theories characteristically postulate spin-dependent interactions for which the ratio of the coupling constants to electrons and nucleons have no relationship to electron and nucleon gyromagnetic ratios. It is also of interest to note that the exotic field may have a transient \cite{Pos13,Pus13} or oscillating \cite{Bud14} time-dependent behavior.

All magnetic shielding materials used in present experiments are based on the interaction of electrons with magnetic fields. If the exotic spin-dependent interaction couples primarily to nuclear spin, the magnetic response of the shield will be negligible. Thus the sensitivity of an experiment to purely nuclear spin-dependent interactions is essentially unaffected by the presence of magnetic shields.

In the case of an exotic interaction coupling to electron spins or electron orbital angular momentum, it is conceivable that a magnetic shield could be affected by the interaction. For simplicity, we assume in this work, as do many theories postulating new interactions \cite{Moo84,Dob06,Fla09,Gra13,Ark05,Geo07,Lia07}, that there is no coupling of the exotic field to orbital angular momentum $\mb{L}$. In the context of quantum field theory, this theoretical bias can be understood as follows. If an exotic field couples to $\mb{L}$, that implies that the field couples to particle current. However, the lowest-order coupling to particle current vanishes if the exotic interaction is mediated by a spin-0 particle (such as an axion or axion-like particle \cite{Moo84,Gra13}). On the other hand, a coupling of a generic massive spin-1 boson to particle current is forbidden by gauge invariance \cite{Dob05}, and constraints on couplings of massless spin-1 bosons are already quite stringent \cite{App03}. Thus, generally, couplings of exotic fields to particle current, and thus $\mb{L}$, are expected to be suppressed relative to spin couplings. Nonetheless, it should also be noted that there are theories that do postulate exotic couplings to $\mb{L}$. For example, hidden photons can mix with ordinary photons, and thus can produce real magnetic fields in magnetically shielded regions that would indeed couple to $\mb{L}$ \cite{Cha15}.  At any rate, in this work we consider an exotic field $\bs{\Upsilon}$ that generates a coupling only to intrinsic spin of the form
\begin{align}
\sH = \xi \bs{\Upsilon} \cdot \mb{S}~,
\label{Eq:exotic-electron-spin-coupling-Hamiltonian}
\end{align}
where $\xi$ is a dimensionless coupling constant for electrons to $\bs{\Upsilon}$ and $\mb{S}$ is the electron spin.

\begin{figure}
\center
\includegraphics[width=3.0 in]{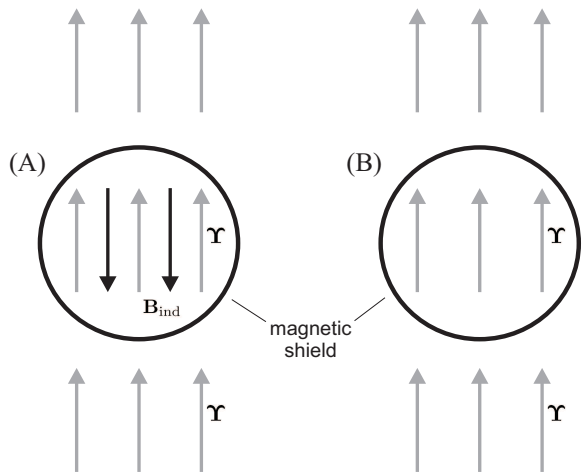}
\caption{Schematic diagram of possible effects of an exotic field $\bs{\Upsilon}$ on a magnetic shield: (A) the shield material may interact with $\bs{\Upsilon}$ to generate an induced field $\mb{B}\ts{ind}$ that approximately cancels the effect of $\bs{\Upsilon}$ on electron spins, or (B) the shield material may have no response to $\bs{\Upsilon}$.}
\label{Fig:exotic-field-shielding-schematic}
\end{figure}

Another important point is that $\bs{\Upsilon}$ is not {\emph{shielded}} per se as magnetic fields are, but rather the primary observable effect of the interaction between the shield and $\bs{\Upsilon}$ is the generation of an induced magnetic field $\mb{B}\ts{ind} \propto -\xi\bs{\Upsilon}$ due to magnetization of the shield by $\bs{\Upsilon}$. This is because the spin-dependent energy shift from an induced exotic field, $\bs{\Upsilon}\ts{ind} \propto \xi\bs{\Upsilon}$, on electron spins would be proportional to $\xi^2$, whereas the spin-dependent energy shift from $\mb{B}\ts{ind}$ is linear in $\xi$ and therefore considerably larger since it is assumed that $\xi \ll 1$. Furthermore, since $|\bs{\Upsilon}\ts{ind}| \ll |\bs{\Upsilon}|$, we can approximate that the response of samples in the interior of the shield is dominated by only $\bs{\Upsilon}$ and $\mb{B}\ts{ind}$ (Fig.~\ref{Fig:exotic-field-shielding-schematic}).

One can imagine two possibilities for the response of a magnetic shield to $\bs{\Upsilon}$: (A) the shield generates an induced field
\begin{align}
\mb{B}\ts{ind} \approx -\frac{\xi}{g\mu_B}\bs{\Upsilon}~,
\label{Eq:B-ind}
\end{align}
where $g$ is the electron $g$-factor, that would approximately cancel the response of electrons to $\bs{\Upsilon}$ within the shield, or (B) the shield has little or no response to $\bs{\Upsilon}$ so that $\mb{B}\ts{ind} \approx 0$, in which case the experimental sensitivity to exotic spin-dependent interactions would be unaffected. (Note that here we are considering the field in a region within a hollow magnetic shield, see Fig.~\ref{Fig:exotic-field-shielding-schematic}, where the magnetization $\mb{M} = 0$, so $\mb{B}\ts{ind} = \mb{H}\ts{ind}$.)

In case (A), if the ensemble of spin-polarized particles used to search for the exotic coupling consists only of electrons, the approximate cancelation between the effect of the exotic interaction and the induced magnetic field $\mb{B}\ts{ind}$ would significantly reduce the size of any observable effect. However, nearly all recent experiments employ some version of the technique of comagnetometry, where the response of multiple species to $\bs{\Upsilon}$ is simultaneously measured. Suppose, for example, that an experiment employs both an ensemble of polarized electrons and an ensemble of polarized neutrons. Within the shield both $\bs{\Upsilon}$ and $\mb{B}\ts{ind}$ are present, and so the polarized electrons do not respond. However, if we assume no coupling of $\bs{\Upsilon}$ to neutron spin, the neutrons will respond to the magnetic torque generated by $\mb{B}\ts{ind}$ and could produce a detectable effect. This highlights the critical importance of both comagnetometry and understanding the response of magnetic shielding in the search for exotic spin couplings.

A further consideration is that most experiments searching for exotic spin-dependent interactions employ multi-layered magnetic shields. Fortunately, this does not change the basic concepts of our arguments because of the superposition principle. Consider, for example, a two-layer shield. According to our model, in case (A), $\bs{\Upsilon}$ essentially acts equivalently to a weak external magnetic field on the outer shield layer. This magnetizes the outer shield layer and produces an induced field $\mb{B}_1$ according to Eq.~\eqref{Eq:B-ind}. The inner shield layer has magnetization produced by the combined effect of $\mb{B}_1$ and $\bs{\Upsilon}$. The superposition principle can be used to conceptually separate the effect of $\mb{B}_1$ and $\bs{\Upsilon}$ on the inner shield layer: according to this perspective within the interior volume of the inner shield layer there is an induced field $\mb{B}_2$ generated by $\bs{\Upsilon}$ according to Eq.~\eqref{Eq:B-ind} and a residual attenuated field $\mb{B}_1^*$ arising from imperfect shielding of $\mb{B}_1$ by the inner layer. Since the magnitude of $\mb{B}_2$ is much greater than the magnitude of $\mb{B}_1^*$, to first order $\mb{B}_1^*$ can be neglected and we can conclude that a multi-layer shield of type (A) has effectively the same behavior as a single-layer shield with respect to exotic interactions. (In fact, $\mb{B}_2$ ends up being equal to the induced field in the case where there is no outer shield layer.)

Considering the effect of exotic spin-dependent interactions on magnetic shielding is particularly important when interpreting the results of null experiments as constraints on new physics: does the shielding reduce or enhance the sensitivity of the experiment to electron spin couplings? Do searches for exotic nuclear spin couplings also constrain electron spin couplings because of the effect of the magnetic shielding illustrated in case (A) above? In the next section we explore the physics of different magnetic shielding materials to determine their expected response to exotic spin-dependent interactions.

\section{Characteristics of different types of magnetic shielding}

\subsection{Superconducting shields}

Superconducting shields are based on the Meissner effect \cite{Mei33}: below the superconducting transition temperature, screening currents are induced that cancel external magnetic fields in the interior of the superconductor (beyond the London penetration depth). The Meissner effect relies on the coupling of magnetic fields to the motion of charged particles rather than the coupling of magnetic fields to the intrinsic spins of electrons. Thus, since the Meissner effect is unrelated to interactions with the electron spins and since we have assumed that the exotic field $\bs{\Upsilon}$ does not couple to orbital angular momentum, $\bs{\Upsilon}$ would produce no screening currents. Therefore the use of superconducting shields would have no effect on experimental sensitivity to $\bs{\Upsilon}$. Hence superconducting shields are an example of case (B) discussed above (Fig.~\ref{Fig:exotic-field-shielding-schematic}).

\subsection{Shielding of ac fields by induction}

Ordinary electromagnetic induction will act to prevent change to the magnetic flux passing through a conductor. Thus ac fields can be shielded due to electric currents induced in conductors. As in the case of the Meissner effect, this shielding mechanism originates from the induced motion of charges in the conductor, and is not related to the intrinsic spins. Thus a rapidly time-varying $\bs{\Upsilon}$ (for example, an oscillating axion field \cite{Gra13,Bud14}) would not be shielded, and so conducting shields are another example of case (B).

\subsection{Soft ferromagnets and ferrimagnets}

The most commonly used types of shielding materials in experiments searching for exotic spin-dependent interactions (as well as magnetometric studies) are soft ferromagnets such as mu-metal (an alloy consisting of 77\% nickel, 16\% iron, 5\% copper, and 2\% molybdenum or chromium) that come in a variety of brand names and alloy compositions (see, for example, Refs.~\cite{Bud13book} and \cite{Alt15} for detailed discussion of such magnetic shielding). The shielding capability of these materials arises from their high magnetic permeability which provides a low-reluctance path for magnetic flux that steers magnetic field lines around the shielded volume. The high magnetic permeability, in turn, is a consequence of the fact that the material magnetization is easily changed by external fields \cite{Chi97}.  Another soft magnetic material used for shielding external magnetic fields is Mn-Zn ferrite \cite{Kor07}. Ferrites, like soft ferromagnets, can have high magnetic permeability (yet typically smaller than that of the best soft ferromagnetic shielding alloys) but also have the attractive feature of high resistivity (typically many orders of magnitude greater than for soft ferromagnetic alloys) which reduces thermal magnetic field noise due to Johnson currents. Ferrites are ferrimagnetic materials, which consist of populations of atoms with opposing magnetic moments; in the case of ferrites suitable for magnetic shields the opposing magnetic moments are unequal and thus do not cancel, leaving residual spontaneous magnetism that can respond to the external magnetic field. The magnetization of both ferromagnets and ferrimagnets is characterized by domain structure, and thus to understand their response to exotic fields with spin-dependent couplings, we must consider the physical mechanisms through which the domain structure changes.

When the magnetic field $\mb{B}\ts{ext}$ outside the magnetic shield changes, there are two primary mechanisms that can contribute to a change in the shield material's magnetization \cite{Chi97}: (a) domain wall displacement and (b) rotation of the domain magnetization. It is the nature of these two processes that determines whether an exotic field $\bs{\Upsilon}$ can produce appreciable effects leading to an induced field $\mb{B}\ts{ind}$ within the shield interior. In particular, we are interested in the case of small field changes, since based on existing constraints on electron-spin-dependent interactions \cite{Hec08,Hec13}, in many cases of interest $\bs{\Upsilon}$ would correspond to magnetic field strengths that are on the order of $10^{-14}~{\rm G}$ or less, although short-range \cite{Ter15}, transient \cite{Pos13,Pus13}, and oscillating exotic fields \cite{Bud14} may be able to produce effects equivalent to larger magnetic field strengths (albeit still $\ll 10^{-6}~{\rm G}$).

Domains of opposing magnetization, separated by domain walls (regions where the magnetization rotates), naturally form in ferromagnetic and ferrimagnetic materials to minimize magnetostatic energy from the self-field interaction, which is proportional to the domain width. On the other hand, the domain-wall energy per unit volume increases as the domain width decreases, and so the characteristic domain size results from a balance of these competing factors \cite{Chi97}. In real materials, the domain structure is influenced by both the crystalline structure and irregularities (voids, nonmagnetic inclusions, internal stresses, grain boundaries, etc.). The crystalline structure of a ferromagnetic material generates a magnetic anisotropy through a pseudodipolar interaction, a variation in the exchange energy with the rotation of the magnetization relative to so-called ``easy'' and ``hard'' axes. In other words, when the magnetization is aligned with the easy axis the exchange energy is minimized, when the magnetization is aligned with the hard axis the exchange energy is maximized. In the case of soft ferromagnetic materials such as mu-metal, the overall magnetic anisotropy is typically produced during annealing in the presence of an applied magnetic field in a hydrogen atmosphere. The annealing process aligns grains and removes impurities, which also reduces the effects of local anisotropies.

The local magnetization in zero field lies parallel to the easy axis, in either the positive or negative direction. Consider two neighboring domains with oppositely oriented magnetization separated by a domain wall. When a field $\mb{B}\ts{ext}$ is applied in the direction of the magnetization of the first domain, the domain wall moves so as to expand the volume of the first domain at the expense of the second domain. This effect, known as domain-wall displacement, can be modeled as a pressure exerted on domains with magnetization along $\mb{B}\ts{ext}$ to expand \cite{Chi97}. For an ideal homogeneous material, the domain wall energy itself is independent of position (in neutral equilibrium), and therefore domains walls can move freely based only on the pressure exerted from externally applied fields. However, in the practical case, the domain-wall energy turns out to be position-dependent (due to impurities, internal stresses, etc.), and the domain wall position will settle into a local minimum. (In fact, it is usually assumed that domain walls exist at most or all energy minima in the material.) Thus domain walls can become pinned at particular sites and in order for the domain wall to be displaced and expansion or contraction of domains to occur, some energy barrier must be overcome. On the other hand, as pointed out by N\'eel \cite{Nee46}, domain walls are also deformable -- so that even though parts of the wall may be pinned at particular sites in the crystal, in general the wall can bulge under magnetic pressure. When the magnetic pressure becomes strong enough to drive the domain wall out of the stable, pinned configuration, the wall displaces irreversibly. The discrete jumping of domain walls from pinned site to pinned site is one of the causes of Barkhausen noise in ferromagnetic and ferrimagnetic materials \cite{Bar19}. In summary, for sufficiently small external magnetic fields, the magnetization of the ferromagnetic material changes via bulging and contracting of domains, while for larger magnetic fields there are irreversible jumps of domain wall positions between pinning sites.

The important point relevant to our considerations is that a continuous, reversible mechanism for domain expansion and contraction, and thus change in magnetization, exists for soft ferromagnets and ferrimagnets: namely, the deformation of the domain walls. Although the magnetic permeability of a soft ferromagnet or ferrimagnet depends on the strength of the external field being shielded for field strengths above some threshhold, for sufficiently small external field strengths where domain wall deformation rather than displacement dominates, the permeability is expected to become relatively independent of the field strength in contrast with the expectations of Refs.~\cite{Rit90,Pan92,Chu93}.

In other soft ferromagnetic materials such as Isoperm (an alloy consisting of 50\% nickel and 50\% iron) and Permalloy (an alloy consisting of 80\% nickel and 20\% iron) where the crystal is strongly stressed, the dominant physical mechanism accounting for the high permeability is rotation and reversal of domain magnetization rather than domain wall displacement and deformation. Nonetheless, there is a similar dichotomy of physical mechanisms: for sufficiently large changes in the external field there can be spontaneous reversal or realignment of domain magnetization, connected to Barkhausen noise \cite{Bar19}, while for smaller external field changes there is continuous and reversible change of domain magnetization through rotation of the magnetization away from the easy axis \cite{Chi97}.

When the external field is near zero, the magnetization behavior of soft ferromagnets and ferrimagnets can be described by
\begin{align}
\mb{M} \approx \prn{ \chi_0 + \frac{\eta}{2} |\mb{H}| } \mb{H}~,
\label{Eq:Rayleigh-loops}
\end{align}
where $\mb{M}$ is the magnetization, $\chi_0$ is the zero-field susceptibility, $\mb{H}$ is the external field, and $\eta$ is the Rayleigh constant \cite{Chi97}. If $|\mb{H}|$ is sufficiently small, the first term, $\chi_0 \mb{H}$ (related to the continuous and reversible processes of domain deformation and magnetization rotation), dominates and the magnetization exhibits a linear dependence on the external field. For larger values of $|\mb{H}|$, the second term, $\eta |\mb{H}|\mb{H}/2$ (related to irreversible magnetization processes such as discrete jumps of pinned domain walls and magnetization reversal), becomes important. Depending on whether the external field increases or decreases the sign of $\eta$ changes, and $\mb{M}$ exhibits hysteresis as a function of changing $\mb{H}$, leading to so-called ``Rayleigh loops'' in the dependence of $\mb{M}$ on $\mb{H}$. In general, the characteristics of the minor hysteresis loops depend on the initial value of $\mb{H}$ and the locus of $\mb{H}$ as it varies, and so the effective permeability can be different depending on the magnetic conditions and history, as studied for mu-metal shielding in Ref.~\cite{Yam05}. To give a rough idea of the scale of these effects, estimates and measurements indicate that the nonlinear term in Eq.~\eqref{Eq:Rayleigh-loops}, due to irreversible processes in mu-metal, can be neglected for external fields $\mb{B}\ts{ext} \lesssim 10^{-3}~{\rm G}$ \cite{Yam05}.

Broadly speaking, for exotic fields $\bs{\Upsilon}$ that couple to electron spin [Eq.~\eqref{Eq:exotic-electron-spin-coupling-Hamiltonian}], soft ferromagnetic and ferrimangetic shields are thus expected to respond and generate an induced field $\mb{B}\ts{ind} \approx -\xi \bs{\Upsilon}/(g \mu_B)$ in the interior of the shield volume that will approximately compensate the effect of $\bs{\Upsilon}$ on electron spins, corresponding to case (A) described above (Fig.~\ref{Fig:exotic-field-shielding-schematic}).

\begin{figure}
\center
\includegraphics[width=3.0 in]{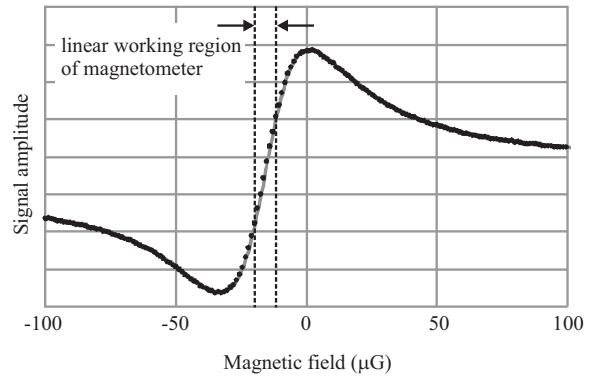}
\caption{Scan of the the applied magnetic field within the innermost magnetic shield layer and measured amplitude of nonlinear magneto-optical rotation with frequency modulated light (FM NMOR), centered on the zero-field ($n=0$) resonance \cite{Bud02,Kim09}. The working region of the magnetometer (width $\approx 6~{\rm \mu G}$), where the signal is linear in response to the applied field, is designated by the vertical dashed lines. The center of the resonance is offset from zero due to a residual $\sim 20~{\rm \mu G}$ field from the innermost shield.}
\label{Fig:FM-NMOR-resonance}
\end{figure}

To test the permeability of mu-metal for relatively small external magnetic fields in the range $10^{-6}~{\rm G} \lesssim B\ts{ext} \lesssim 10^{-3}~{\rm G}$, below the level where the effect of Rayleigh loops should be negligible, we carried out a measurement of the field inside the innermost layer of a five-layer mu-metal magnetic shield when an external field $\mb{B}\ts{ext}$ was generated by a current through a coil wrapped around the same innermost shield layer. The experimental setup is in most respects identical to the setup used to study nonlinear magneto-optical rotation with frequency modulated light (FM NMOR) described in Ref.~\cite{Kim09}. The magnetic field inside the innermost shield is measured using the FM NMOR technique \cite{Kim09,Bud02}: linearly polarized laser light, near-resonant with an atomic transition, propagates through an atomic vapor cell along the direction of the magnetic field to be measured; the laser light is frequency modulated at $\Omega_m \approx 2\pi \times 1000~{\rm Hz}$ (with a modulation depth of $\approx 100~{\rm MHz}$) and optical rotation of the laser light is measured at the output with a lock-in amplifier referenced to the laser frequency modulation at $\Omega_m$. Prominent resonances in the magnetic field dependence of the optical rotation amplitude measured at the first harmonic of $\Omega_m$ are observed when
\begin{align}
n \Omega_m = 2 \Omega_L~,
\label{Eq:FM-NMOR-resonance-condition}
\end{align}
where $\Omega_L$ is the Larmor frequency and $n=0,1$. In this study, we employ the $n=0$ resonance to measure near-zero magnetic fields within the innermost shield, which means that we work at near-zero magnetic field. For our experiment, the laser light power is $\approx 150~{\rm \mu W}$ and the laser is detuned $\approx 400~{\rm MHz}$ to the high-frequency wing of the Doppler-broadened $^{85}$Rb $F=3 \rightarrow F'$ hyperfine component of the D2 transition to maximize the signal (see Ref.~\cite{Kim09} for more details).

\begin{figure}
\center
\includegraphics[width=3.0 in]{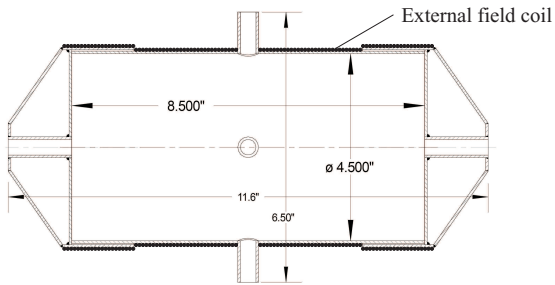}
\caption{The innermost mu-metal magnetic shield and schematic of external field coil.}
\label{Fig:innermost-shield-schematic}
\end{figure}

A 192-turn coil (the ``external field'' coil) is wrapped around the outside of the innermost shield layer (Fig.~\ref{Fig:innermost-shield-schematic}). A 5-Hz sinusoidally oscillating current generated with a Model 200CD Hewlett-Packard oscillator passes through the external field coil. The oscillating magnetic field within the innermost shield layer observed with FM NMOR is measured with another lock-in amplifier. Throughout these measurements, the magnetic fields measured are sufficiently small so that the $n=0$ FM NMOR signal is linearly proportional to the magnetic field. The shielding factor is estimated as the ratio of expected magnetic field on the cell calculated in the absence of the innermost shield layer to the actually measured magnetic field within the innermost shield layer.

\begin{figure}
\center
\includegraphics[width=3.0 in]{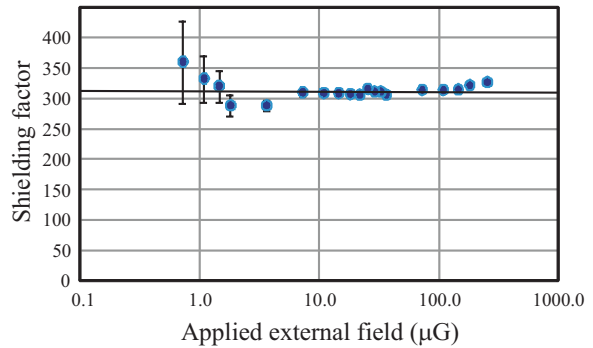}
\caption{Estimated shielding factor (ratio of calculated field without innermost shield to measured magnetic field with innermost shield) as a function of the applied external field. For data with larger external field magnitudes, see Ref.~\cite{Yam05}; the upper limit of the field range was defined by the linear working region of the magnetometer (see Fig.~\ref{Fig:FM-NMOR-resonance}).}
\label{Fig:shielding-factor-vs-external-field}
\end{figure}

The measured shielding factor as a function of applied external field is shown in Fig.~\ref{Fig:shielding-factor-vs-external-field}, demonstrating that over the range $10^{-6}~{\rm G} \lesssim B\ts{ext} \lesssim 10^{-3}~{\rm G}$ the shielding factor for mu-metal is relatively constant, as predicted from the fact that the contribution from Rayleigh loops is negligible in this field range. This supports the conclusion that soft ferromagnetic or ferrimagnetic shielding indeed responds to small field variations and the suggestion of Refs.~\cite{Vor88,Bob91} that soft ferromagnets can indeed be affected by exotic spin-dependent interactions. As noted in the introduction, the lower limit of the studied field range is particularly relevant since it is the usual range of the ambient magnetic fields acting on the innermost shielding layer in experiments, and torques from exotic fields would sum with the magnetic torques studied here.

\section{Flux concentrators and exotic fields}

\begin{figure}
\center
\includegraphics[width=3.3 in]{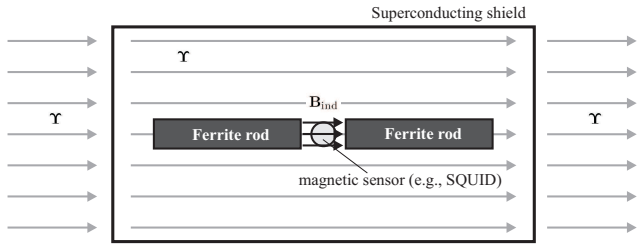}
\caption{Schematic diagram (not to scale) of a flux concentrator applied to a search for an exotic field $\Upsilon$ that couples to electron spins. The magnetic field induced by $\Upsilon$ in the gap between the ferrite rods, depending on the choice of ferrite material and the exact geometry, can generate an energy shift of electron spin states over an order of magnitude greater than $\Upsilon$ by itself. The induced magnetic field $\mb{B}\ts{ind}$ can be measured with a magnetic sensor such as a SQUID (Superconducting QUantum Interference Device).}
\label{Fig:flux-concentrator-setup}
\end{figure}

The above analysis raises the possibility of amplifying the effect of an exotic field $\Upsilon$ coupling to electron spins by using a flux concentrator to generate an induced field $\mb{B}\ts{ind}$ proportional to $\Upsilon$ that produces a larger effect on a detector than $\Upsilon$ by itself would. Consider, for example, the arrangement shown in Fig.~\ref{Fig:flux-concentrator-setup}. The flux concentrator and magnetic sensor can be placed within a superconducting shield to block external magnetic field variations. As discussed above, the superconducting shield would have no effect on $\Upsilon$. A flux concentrator made of, for example, a ferrite material would respond to $\Upsilon$ by generating an internal magnetization as the electron spins respond to $\Upsilon$ as they would to a weak magnetic field. As opposed to the case of magnetic shielding discussed above, the induced magnetic field $\mb{B}\ts{ind}$ in the gap between the ferrite rods would lead to a larger response as compared to $\Upsilon$ alone, namely:
\begin{align}
\mb{B}\ts{ind} \approx G\frac{\xi}{g\mu_B}\bs{\Upsilon}~,
\end{align}
where $G$ is the flux concentration factor. In the case of a similar geometry employing ferrite rods of length $\approx 5~{\rm cm}$, diameter $\approx 3~{\rm mm}$, and separation $\approx 2~{\rm mm}$, $G \approx 20$ was obtained \cite{Gri09} (even higher values for $G$ may be achieved with triangular or conic geometries). Ferrite, being an insulator, has the attractive feature of negligible magnetic field noise from thermal currents. However, if the entire setup is designed to work at cryogenic temperatures for which the permeability of ferrites is significantly reduced, a material such as Metglas (an amorphous ferromagnetic material) with good cryogenic properties \cite{Qua04} may be more suitable for the flux concentrator.

\section{Conclusion}

In conclusion, we have considered the response of magnetic shielding materials to exotic spin-dependent interactions. The only case where the shielding material is predicted to respond to the exotic field is when the exotic field couples to electron spin and the shield is made of a soft ferromagnetic or ferrimagnetic material. In this case, the exotic field is predicted to produce an induced magnetic field within the shield. The effect of this induced field should be considered in experiments searching for such effects. Furthermore, this observation raises the possibility of amplifying exotic electron-spin-dependent effects through use of a flux concentrator.

\section{Acknowledgments}

The authors are grateful to Surjeet Rajendran, Peter Fierlinger, and Blayne Heckel for enlightening discussions and to Valeriy Yashchuk for design of the magnetic shields used in our experiment.  DFJK acknowledges support from the National Science Foundation under grant PHY-1307507 and the Heising-Simons and Simons Foundations; DB acknowledges the support of the DFG Koselleck program and the Heising-Simons and Simons Foundations; SP acknowledges support from the Iuventus Plus Program of the Polish Ministry of Science and Higher Education.


\begin{thebibliography}{99}


\bibitem{ATL12}
T. Abajyan {\it{et al.}} (ATLAS collaboration), Phys.\ Lett.\ B {\textbf{716}}, 1 (2012).

\bibitem{CMS12}
V.\ Khachatryan {\it{et al.}} (CMS collaboration), Phys.\ Lett.\ B {\textbf{716}}, 30 (2012).

\bibitem{Sak65}
A. D. Sakharov, Pisma Zh. Eksp. Teor. Fiz. {\textbf{5}}, 32 (1967) [JETP Lett. {\textbf{5}}, 24 (1967)].

\bibitem{Gav94}
M. B. Gavela, P. Hernandez, J. Orloff, O. Pene and C. Quimbay, Nucl. Phys. B {\textbf{430}}, 382 (1994).

\bibitem{Hue95}
P. Huet and E. Sather, Phys. Rev. D {\textbf{51}}, 379 (1995).

\bibitem{Ber05}
Gianfranco Bertone, Dan Hooper, and Joseph Silk, Physics Reports {\textbf{405}}, 279 (2005).

\bibitem{Per99}
S. Perlmutter {\it et al.} (Supernova Cosmology Project), Astrophys.J. {\textbf{517}}, 565 (1999).

\bibitem{Rie98}
A. G. Riess {\it et al.} (Supernova Search Team), Astron.J. {\textbf{116}}, 1009 (1998).

\bibitem{Wei89}
S. Weinberg, Rev. Mod. Phys. {\textbf{61}}, 1 (1989).

\bibitem{Adl95}
R. J. Adler, B. Casey, and O. C. Jacob, Am. J. of Phys., {\textbf{63}}, 620 (1995).

\bibitem{Ark98}
N. Arkani-Hamed, S. Dimopoulos, and G. R. Dvali, Phys. Lett. B {\textbf{429}}, 263 (1998).

\bibitem{Gra15}
P. W. Graham, D. E. Kaplan, and S. Rajendran, Phys. Rev. Lett. {\textbf{115}}, 221801 (2015).



\bibitem{Moo84}
J. E. Moody and F. Wilczek, Phys. Rev. D {\textbf{30}}, 130 (1984).

\bibitem{Dob06}
B. A. Dobrescu and I. Mocioiu, J. High Energy Phys. {\textbf{11}}, 5 (2006).

\bibitem{Fla09}
V. Flambaum, S. Lambert and M. Pospelov, Phys. Rev. D {\textbf{80}}, 105021 (2009).

\bibitem{Gra13}
P. W. Graham and S. Rajendran, Phys. Rev. D {\textbf{88}}, 035023 (2013).


\bibitem{Hun13}
L. Hunter, J. Gordon, S. Peck, D. Ang, J.-F. Lin, Science {\textbf{339}}, 928 (2013).

\bibitem{Hec08}
B. R. Heckel, E. G. Adelberger, C. E. Cramer, T. S. Cook, S. Schlamminger, and U. Schmidt, Phys. Rev. D {\textbf{78}}, 092006 (2008).

\bibitem{Hec13}
B. R. Heckel, W. A. Terrano, and E. G. Adelberger, Phys. Rev. Lett. {\textbf{111}}, 151802 (2013).

\bibitem{Ter15}
W. A. Terrano, E. G. Adelberger, J. G. Lee, and B. R. Heckel Phys. Rev. Lett. {\textbf{115}}, 201801 (2015).

\bibitem{Vas09}
G. Vasilakis, J. M. Brown, T. W. Kornack, and M. V. Romalis, Phys. Rev. Lett. {\textbf{103}}, 261801 (2009).

\bibitem{Kot15}
S. Kotler, R. Ozeri, and D. F. Jackson Kimball, Phys. Rev. Lett. {\textbf{115}}, 081801 (2015).

\bibitem{Led13}
M. P. Ledbetter, M. V. Romalis, and D. F. Jackson Kimball, Phys. Rev. Lett. {\textbf{110}}, 040402 (2013).

\bibitem{Smi11}
M. Smiciklas, J. M. Brown, L. W. Cheuk, S. J. Smullin, and M. V. Romalis, Phys. Rev. Lett. {\textbf{107}}, 171604 (2011).

\bibitem{Kim13}
D. F. Jackson Kimball, I. Lacey, J. Valdez, J. Swiatlowski, C. Rios, R. Peregrina-Ramirez, C. Montcrieffe, J. Kremer, J. Dudley, and C. Sanchez,  Ann. der Physik {\textbf{525}}, 514 (2013).

\bibitem{Tul13}
K. Tullney, F. Allmendinger, M. Burghoff, W. Heil, S. Karpuk, W. Kilian, S. Knappe-Gr\"uneberg, W. M\"uller, U. Schmidt, A. Schnabel, F. Seifert, Yu. Sobolev, and L. Trahms, Phys. Rev. Lett. {\textbf{111}}, 100801 (2013).

\bibitem{Bul13}
M. Bulatowicz, R. Griffith, M. Larsen, J. Mirijanian, C. B. Fu, E. Smith, W. M. Snow, H. Yan, and T. G. Walker, Phys. Rev. Lett. {\textbf{111}}, 102001 (2013).

\bibitem{Pos13}
M.\ Pospelov, S.\ Pustelny, M.P.\ Ledbetter, D.F.\ Jackson Kimball, W.\ Gawlik, and D.\ Budker, Phys. Rev. Lett. {\textbf{110}}, 021803 (2013).

\bibitem{Pus13}
S. Pustelny, D. F. Jackson Kimball, C. Pankow, M. P. Ledbetter, P. Wlodarczyk, P. Wcislo, M. Pospelov, J. R. Smith, J. Read, W. Gawlik, and D. Budker, Ann. der Physik {\textbf{525}}, 659 (2013).

\bibitem{Bud14}
D. Budker, P. W. Graham, M. Ledbetter, S. Rajendran, and A. O. Sushkov, Phys. Rev. X {\textbf{4}}, 021030 (2014).

\bibitem{Arv14}
A. Arvanitaki and A. A. Geraci, Phys. Rev. Lett. {\textbf{113}}, 161801 (2014).

\bibitem{Ant11}
I. Antoniadis, S. Baessler, M. B\"uchner, V. V. Fedorov, S. Hoedl, A. Lambrechti, V. V. Nesvizhevsky, G. Pignol, K. V. Protasov, S. Reynaud, and Yu. Sobolev, C. R. Physique {\textbf{12}}, 755 (2011).


\bibitem{Bud13book}
D. Budker and D. F. Jackson Kimball, eds., {\it{Optical Magnetometry}} (Cambridge University Press, Cambridge, 2013).


\bibitem{Vor88}
P. V. Vorobyov and Ya. I. Gitarts, Phys. Lett. B {\textbf{208}}, 146 (1988).

\bibitem{Bob91}
V. F. Bobrakov, Yu. V. Borisov, M. S. Lasakov, A. P. Serebrov, R. R. Tal'daev, and A. S. Trofimova, Pis'ma
Zh. Eksp. Teor. Fiz. {\textbf{53}}, 283 (1991) [JETP Lett. {\textbf{53}}, 294 (1991)].

\bibitem{Rit90}
R. C. Ritter, C. E. Goldblum, W.-T. Ni, G. T. Gillies, and C. C. Speake, Phys. Rev. D {\textbf{42}}, 977 (1990).

\bibitem{Pan92}
S.-s. Pan, W.-T. Ni, and S.-C. Chen, Mod. Phys. Lett. A {\textbf{7}}, 1287 (1992).

\bibitem{Chu93}
T. C. P. Chui and W.-T. Ni, Phys. Rev. Lett. {\textbf{71}}, 3247 (1993).




\bibitem{Kim09}
D. F. Jackson Kimball, L. R. Jacome, S. Guttikonda, E. J. Bahr, and L. F. Chan, J. Appl. Phys. {\textbf{106}}, 063113 (2009).



\bibitem{Nev80}
D. E. Neville, Phys. Rev. D {\bf{21}}, 2075 (1980).

\bibitem{Nev82}
D. E. Neville, Phys. Rev. D {\bf{25}}, 573 (1982).

\bibitem{Car94}
S. M. Carroll and G. B. Field, Phys. Rev. D {\bf{50}}(6), 3867 (1994).

\bibitem{Kos11}
 V. A. Kosteleck\'{y} and N. Russell, Rev. Mod. Phys. {\textbf{83}}, 11 (2011).

\bibitem{Ark05}
N. Arkani-Hamed H.-C. Cheng, M. Luty and J. Thaler, J. High Energy Phys. {\textbf{07}}, 029 (2005).

\bibitem{Geo07}
H. Georgi, Phys. Rev. Lett. {\textbf{98}}, 221601 (2007).

\bibitem{Lia07}
Y. Liao and J.Y. Liu, Phys. Rev. Lett. {\textbf{99}}, 191804 (2007).



\bibitem{Dob05}
B. A. Dobrescu, Phys. Rev. Lett. {\textbf{94}}, 151802 (2005).

\bibitem{App03}
T. Appelquist, B. A. Dobrescu, and A. R. Hopper, Phys. Rev. D {\textbf{68}}, 035012 (2003).

\bibitem{Cha15}
S. Chaudhuri, P. W. Graham, K. Irwin, J. Mardon, S. Rajendran, and Y. Zhao, Phys. Rev. D {\textbf{92}}, 075012 (2015).




\bibitem{Mei33}
W. Meissner and R. Ochsenfeld, Naturwissenschaften {\textbf{21}} 787 (1933).



\bibitem{Chi97}
S. Chikazumi, {\it Physics of Ferromagnetism} 2nd edition (Oxford University Press, Oxford, 1997).

\bibitem{Alt15}
I. Altarev, M. Bales, D. H. Beck, T. Chupp, K. Fierlinger, P. Fierlinger, F. Kuchler, T. Lins, M. G. Marino, B. Niessen, G. Petzoldt, U. Schl\"apfer, A. Schnabel, J. T. Singh, R. Stoepler, S. Stuiber, M. Stur, B. Taubenheim, and J. Voigt, J. Appl. Phys. {\textbf{117}}, 183903 (2015).

\bibitem{Kor07}
T. W. Kornack, S. J. Smullin, S.-K. Lee, and M. V. Romalis, Appl. Phys. Lett. {\textbf{90}}, 223501 (2007).


\bibitem{Nee46}
L. N\'eel, Annal. Univ. Grenoble {\textbf{22}}, 299 (1946).

\bibitem{Bar19}
H. Barkhausen, Phys. Z. {\textbf{20}}, 401 (1919).

\bibitem{Yam05}
K. Yamazaki, K. Kato, K. Muramatsu, A. Haga, K. Kobayashi, K. Kamata, K. Fujiwara, and T. Yamaguchi, IEEE Trans. Magnetics {\textbf{41}}, 4087 (2005).

\bibitem{Bud02}
D. Budker, D. F. Kimball,  V. V. Yashchuk, and M. Zolotorev,
Phys. Rev. A {\textbf{65}}, 055403 (2002).

\bibitem{Yas00}
V. V. Yashchuk, D. Budker, and J. Davis, Rev. Sci. Instrum. {\textbf{71}}, 341 (2000).

\bibitem{Bal10}
M. V. Balabas, T. Karaulanov, M. P. Ledbetter, and D. Budker, Phys. Rev. Lett. {\textbf{105}}, 070801 (2010).

\bibitem{Xu06}
S. Xu, S. M. Rochester, V. V. Yashchuk, M. H. Donaldson, and D. Budker, Rev. Sci. Instrum. {\textbf{77}}, 083106 (2006).



\bibitem{Gri09}
W. C. Griffith, R. Jimenez-Martinez, V. Shah, S. Knappe, and J. Kitching, Appl. Phys. Lett. {\textbf{94}}, 023502 (2009).

\bibitem{Qua04}
H. P. Quach and T.C.P Chui, Cryogenics {\textbf{44}}, 445 (2004).

\end{thebibliography}
\end{document}